\documentstyle[preprint,aps]{revtex}

\tightenlines

\begin{document}

\title{Probing Transport Theories via Two-Proton Source Imaging}

\author{G. Verde$^a$, P. Danielewicz$^a$, W.G. Lynch$^a$, D.A. Brown$^b$,
C.K. Gelbke$^a$, M.B. Tsang$^a$}

\author{\small $^a$ National Superconducting Cyclotron Laboratory and
Department of Physics and Astronomy, Michigan State University,
East Lansing, MI 48824, USA}
\author{\small $^b$ Lawrence Livermore National Laboratory, Livermore, CA 94551-0808, USA}

\date{\today}

\maketitle

\begin{abstract}
Imaging technique is applied to two-proton correlation functions
to extract quantitative information about the space-time
properties of the emitting source and about the fraction of
protons that can be attributed to fast emission mechanisms. These
new analysis techniques resolve important ambiguities that
bedeviled prior comparisons between measured correlation functions
and those calculated by transport theory. Quantitative comparisons
to transport theory are presented here. The results of the present
analysis differ from those reported previously for the same
reaction systems. The shape of the two-proton emitting sources are
strongly sensitive to the details about the in-medium
nucleon-nucleon cross sections and their density dependence.
\end{abstract}


{\bf INTRODUCTION}

Transport theories have been extensively used to describe the main
features of heavy ion collisions at intermediate energies
\cite{Ber88,Aic91,Ono92,Ono96}. Successful microscopic models have
been based on the Boltzmann-Uehling-Uhlenbeck (BUU) equation,
which describes the temporal evolution of the one-body phase-space
density under the influence of the nuclear mean field and
individual nucleon-nucleon collisions \cite{Ber88}. The importance
of such models stems from the connections they provide between
observables measured in nucleus-nucleus collisions and microscopic
quantities like the nuclear mean fields and the in-medium cross
section. Recently, this connection has been exploited to place
constraints on the equation of state (EOS) of nuclear matter at
densities of $2\rho_{0}\leq\rho\leq 4\rho_{0}$, where $\rho_0$ is
the nuclear saturation density \cite{Dannew}. Comparable
constraints on the EOS at lower densities require a more detailed
understanding of transport phenomena at intermediate energies
where the delicate interplay of competing sources of pressure,
such as collisions via the residual interaction, govern the
collision dynamics \cite{Mag00}.

Two-proton correlation functions can provide an important test of
transport theory \cite{Gon90a,Kun93,Han95}, through their
sensitivity to the space-time properties of nuclear reactions
\cite{Koo77,Han95,Hei99}. Initial applications of BUU transport
theory to two-proton correlation functions at incident energies of
E/A$<$100 MeV were successful and indicated a significant
sensitivity of the calculated correlation functions to the
in-medium cross section \cite{Gon90a,Kun93,Han95}. The application
of such techniques to higher incident energies, however, revealed
that there were significant problems in reconciling the stronger
calculated correlation functions to the weaker measured ones
\cite{Han95}. These problems, discussed below, led to criticisms
that BUU transport theory might be an inadequate theoretical tool
for such studies, either due to the neglect of the many body
correlations in the BUU approach \cite{Kun93} or due to the
neglect of the long lived decays of unstable fragments emitted
during a collision \cite{Han95}.

In this paper, we largely resolve this issue by showing how more
quantitative experimental analyses of two-proton correlations can
provide information that can be quantitatively compared to BUU
transport theory. We begin by reviewing the basic Koonin-Pratt
theoretical approach \cite{Koo77,Gon91a} for proton-proton
correlation function analyses. We discuss the basic assumptions
and limitations of previous analyses that typically assumed a
single emission mechanism that could be approximated by one source
of radius $r_{0}$ and lifetime $T$ or a dynamical model such as
the BUU model and focused upon the height of the correlation
function maximum at proton relative momenta of 20 MeV/c. We
describe the more quantitative imaging approach that utilizes all
the information contained in the shape of the correlation function
\cite{Ver02,Bro01,Bro97,Bro98}. We apply this approach to
two-proton correlation functions measured in central Ar+Sc
collisions at E/A=80, 120 and 160 MeV where the original problems
were observed \cite{Han95}. With this technique, we determine the
contributions of slow proton emitting sources, allowing direct
comparisons to correlations calculated using the BUU equation. We
use this information to place constraints on the in-medium
nucleon-nucleon cross section.

\vspace{1cm}

{\bf TWO-PROTON CORRELATION FUNCTIONS}

Experimentally, the (angle-averaged) two-proton correlation
function, $1+R(q)$, is defined in terms of the two-particle
coincidence yield, $Y(\vec{p}_{1},\vec{p}_{2})$, and the single
particle yields $Y(\vec{p}_{1})$ and $Y(\vec{p}_{2})$:
\begin{equation}
{\Sigma }Y(\vec{p}_{1},\vec{p}_{2})=k\cdot
\left[1+R(q)\right]\cdot
{\Sigma }\left[ Y(\vec{p%
}_{1})\cdot Y(\vec{p}_{2})\right], \label{eq01}
\end{equation}
Here, $\vec{p}_{1}$ and $\vec{p}_{2}$  are the laboratory momenta
of the coincident particles, $q$ is the momentum of relative
motion, and $C$ is a normalization constant chosen such that
$\langle R(q)\rangle$=0 for large $q$-values where final-state
interaction effects are negligible. For a given experimental
gating condition, the sums on each sides of Eq. (1) extend over
all particle energies and detector combinations corresponding to
each bin of $q$.

Eq. 1 is used to build two-proton correlation functions measured
in central Ar+Sc reactions and shown in Fig. 1. The data points
refer to incident energies of E/A=80 (upper panels), 120 (middle
panels) and 160 MeV (lower panels). Experimental details can be
found in Ref. \cite{Han95} and the references therein. The
different data symbols refer to proton pairs with total momenta,
$P$, in the CM of the reaction, 200$\leq P\leq$ 400 MeV/c
(triangles) and 400$\leq P\leq$ 800 MeV/c (circles). The data are
the same in both the left and right panels.

The shape of these correlation functions reflects the interplay of
the short-range nuclear interaction, the antisymmetrization, and
the long-range Coulomb interaction between the emitted protons
\cite{Koo77,Boa90}. The attractive S-wave nuclear interaction
leads to the observed pronounced maximum at relative momentum
$q\approx$20 MeV/c. At E/A=80 and 120 MeV, it is observed that the
height of the peak at 20 MeV/c is larger for proton pairs with
larger total momenta in the CM of the reaction. This total
momentum dependence nearly disappears at the highest incident
energy, E/A=160 MeV.

Theoretically, the correlation function can be calculated using
the angle-averaged Koonin-Pratt equation \cite{Koo77,Boa90}
\begin{equation}
R(q)=4\pi\int drr^{2}K(q,r)S(r) \label{eq02}
\end{equation}
The angle-averaged kernel $K(q,r)$ is calculated from the radial
part of the antisymmetrical two-proton relative wave-function. At
short distances, the kernel is dominated by the strongly
attractive single S-wave proton-proton interaction which gives
rise to a maximum in $R(q)$ at 20 MeV/c. The long-range repulsive
Coulomb interaction produces a minimum at $q\approx$0 MeV/c to
which the antisymmetrization contributes too. The source function
$S(r)$ is defined as the probability distribution for emitting a
pair of protons with relative distance $r$ at the time the second
proton is emitted. The two-particle source function can be
determined from
\begin{equation}
S_{\vec{P}}(r)=\frac{ \int d^{3}R\cdot
f\left(\frac{1}{2}\vec{P},\vec{R}+\frac{1}{2}\vec{r},t_{>}\right)
f\left(\frac{1}{2}\vec{P},\vec{R}-\frac{1}{2}\vec{r},t_{>}\right)
} { |\int d^{3}r\cdot
f\left(\frac{1}{2}\vec{P},\vec{r},t_{>}\right)|^{2} } \label{eq03}
\end{equation}
where $\vec{R}=\frac{1}{2}\left(\vec{r}_{1}+\vec{r}_{2}\right)$ is
the center of mass coordinate of the two particles and
$\vec{P}=\vec{p}_{1}+\vec{p}_{2}$ is their total momentum. The
Wigner function $f\left(\vec{p},\vec{r},t_{>}\right)$ is the
phase-space distribution of particles with momentum $\vec{p}$  and
position $\vec{r}$ at some time $t_{>}$ after both particles have
been emitted. The function $f\left(\vec{p},\vec{r},t_{>}\right)$
can be expressed in terms of the single particle emission function
$d\left(\vec{p},\vec{r},t\right)$, i.e., the distribution of last
emission points for a particle with momentum $\vec{p}$ at location
$\vec{r}$ and time $t$ \cite{Gon91a}:
\begin{equation}
f\left(\vec{p},\vec{r},t_{>}\right)=\int_{-\infty}^{t_{>}} dt\cdot
d\left(\vec{p},\vec{r}-\vec{p}\left(t_{>}-t\right)/m,t\right)
\label{eq04}
\end{equation}

From Eqs. (2-4), it can be seen that the two-proton correlation
function is directly related to the space-time extent of the
source that emits the protons. For fast emission where the source
spatial dimension exceeds the product of the proton velocity times
the source lifetime, the spatial properties dominate the source
function. Such is the case for the pre-equilibrium emission
predicted by transport models of nuclear reactions where emission
timescales are often less than 100 fm/c. For slow emission, on the
other hand, the lifetime results in an extended source
\cite{Ver02,Gon92,Dey89,Dey90,Gon90b,Gou91,Elm91,Gon91b,Lis94}.
Such slow emission processes are characteristic of slow
evaporative proton emission from heavy residua and of the decay of
excited fragments $A>3$, which are produced abundantly in central
collisions, and can decay over very long timescales (of the order
of thousands of fm/c).

The sensitivity of correlation functions to such long time scale
emissions can be small, especially when the source also contains
contributions from short-lived dynamical processes \cite{Ver02}.
Long-lived emitting sources contribute to the source function S(r)
in Eq. (3) mainly at large relative distance values, $r>$10 fm,
where $K(q,r)$ is dominated by the Coulomb interaction. These
large distances influence the correlation function only at very
small relative momenta, $q<$10 MeV/c, making $1+R(q)$ small where
the measurements are difficult to perform
\cite{Koo77,Ver02,Kun90}. In cases where only long timescale
emissions are present, there can also be measurable effects at
larger relative momenta 10 MeV/c$\leq q\leq$20 MeV/c \cite{Lis94},
but these effects are overwhelmed by the contributions to the
correlation function maximum at q=20 MeV/c from pre-equilibrium
emission when such fast emissions are also present \cite{Ver02}.
These effects from long-lived decays are so negligible at relative
momenta $q>$10 MeV/c that quantitative comparisons to fast
emissions predicted by transport theory may therefore be
undertaken provided corrections for the long-lived decays are made
\cite{Han95,Ver02,Kun90}. In order to understand these corrections
we consider the simple limit wherein fast emission provides a
fraction $f$ of the emitted protons, $Y_{1,fast} = f\cdot Y_{1}$,
and the rest of the emission with yield, $Y_{1,slow} = (1- f)\cdot
Y_{1}$, is so slow that it contribute negligibly at relative
momenta $q>$10 MeV/c. In this limit, the correlation function R(q)
will reflect only the fast source and will be given by
\begin{equation}
R(q)=f^{2}\cdot R_{fast}(q)\equiv \lambda\cdot R_{fast}(q)
\label{eq05}
\end{equation}
where $R_{fast}(q)$ denotes the correlation function when only
fast emission is present \cite{Ver02}. Equation (5) stipulates
that the height of the correlation function peak, $R(20 MeV/c)$,
is attenuated by the factor $\lambda=f^{2}$. Thus, in such limit,
the shape of the correlation function peak reflects uniquely the
fast two-proton source function and can be used to measure the
source shape. The height does not provide unambiguous information
about the shape of the fast source, but combined with information
obtained from the shape of the correlation function, provides a
measure of the relative contribution, $f$, of the fast source
\cite{Ver02}. It is this new understanding of two-proton
correlations that enables us to remove the ambiguities that
bedeviled the previous attempts to relate correlation functions to
transport theory.

\vspace{1cm}

{\bf TWO-PROTON CORRELATIONS IN BUU SIMULATIONS}

The Boltzmann-Uheling-Uhlenbeck (BUU) equation \cite{Ber88}
describes the temporal evolution of the one-body density
distribution, $f\left(\vec{p},\vec{r},t_{>}\right)$:
\begin{eqnarray}
\frac{\partial f}{\partial t} &+&
\frac{\vec{p}}{m}\cdot\vec{\nabla}_{r}f\left(\vec{p},\vec{r},t\right)-
\vec{\nabla}_{r}U(\vec{r})\cdot\vec{\nabla}_{p}f\left(\vec{p},\vec{r},t\right)
\nonumber \\
&=& \frac{1}{2\pi^{3}m^{2}}\int d^{3}\vec{q}_{2}\int
d\Omega^{'}\frac{d\sigma}{d\Omega^{'}}
\left[(1-f_{1})(1-f_{2})f_{1}^{'}f_{2}^{'}+(1-f_{1}^{'})(1-f_{2}^{'})f_{1}f_{2}\right]
\label{eq06}
\end{eqnarray}
The terms on the l.h.s. of Eq. (6) account for the changes of
$f\left(\vec{p},\vec{r},t\right)$ due to the motion of particles
in the average mean field potential, $U(\vec{r})$, produced by
other particles. The r.h.s. accounts for changes of
$f\left(\vec{p},\vec{r},t\right)$ due to the nucleon-nucleon
collisions. The single particle distribution
$f\left(\vec{p},\vec{r},t\right)$  calculated from BUU can be
substituted in Eq. (2-4) to calculate the theoretical two-proton
correlation functions for comparison to the experimental data.
This approach has been used to test transport theories by
comparisons with two-proton correlations
\cite{Gon90a,Kun93,Han95}.

One may use such comparisons to constrain the NN collision cross
section $d\sigma/d\Omega^{'}$ and, possibly, the potential
$U(\vec{r})$ in the l.h.s. of Eq. (6). In our calculations we
explored different equations of state. Negligible dependence of
calculated correlation functions on the equation of state was
found at the explored incident energies, in agreement with
previous studies \cite{Gon90a,Han95}. Therefore we have primarily
utilized the stiff EOS with no momentum dependence in the mean
fields to minimize computation times. On the other hand, we have
observed a significant sensitivity to the in-medium cross section.
To explore this sensitivity, we used density dependent in-medium
NN cross section of the form \cite{Dansig,Dan02}
\begin{equation}
\sigma_{in,{\eta}}(\rho)=\eta\cdot\rho^{-2/3}\cdot\tanh\left[\frac{\sigma_{free}}{\eta\cdot\rho^{-2/3}}\right]
\label{eq07}
\end{equation}
where $\sigma_{free}$ is the free NN collision cross section and
$\rho$ is the surrounding nucleon density. Such parametrization of
the density dependence of the in-medium cross sections has been
introduced \cite{Dansig,Dan02} to take into account the fact that
the geometric cross section radius should not exceed the
interparticle distance, $\sigma\lesssim {\eta}{\rho^{-2/3}}$. Eq.
7 has been shown to be successful in describing stopping
observables \cite{Dansig,Dan02}.
Our simulations were performed
for the three different values of $\eta$ listed in Table I. The
case of $\eta=\infty$, labelled $free$, describes simulations with
free NN cross sections (Note: $\lim_{\eta \to \infty}
\sigma_{in,{\eta}}=\sigma_{free}$). Two choices of reduced
in-medium cross sections were used, $\sigma_{in,1}(\rho)$ and
$\sigma_{in,0.6}(\rho)$ corresponding to $\eta$=1.0 and
$\eta$=0.6, respectively. As shown in Table I,
$\sigma_{in,1}(\rho)\approx 0.8\sigma_{free}$ and
$\sigma_{in,0.6}(\rho)\approx 0.6\sigma_{free}$ for values of
$\sigma_{free}=30 mb$ and $\rho=0.17 fm^{-3}$. On the other hand,
$\sigma_{in,{\eta}}=\sigma_{free}$ by construction, when the
density is very low, regardless of the value of $\eta$.

The solid lines on the left panels of Fig. 1 show the results of
BUU simulations for a stiff equation of state (K=380 MeV) with no
momentum dependent interaction and using a free nucleon-nucleon
cross sections ($\eta\to\infty$). The dashed and dotted lines on
the right panels of Fig. 1 show BUU simulated correlation
functions using reduced in-medium cross sections corresponding to
$\sigma_{in,1}$ and $\sigma_{in,0.6}$, respectively. It is evident
that this simplest application of BUU is not able to reproduce the
variations in the height of the peak at 20 MeV/c with incident
energy and proton total momenta. For example, the correlation
functions for high total momenta proton pairs, 400$\leq P\leq$ 800
MeV/c, are systematically overpredicted by the BUU model.
Discrepancies of this magnitude cannot be remedied by making
alternative choices for the EOS, by using momentum-dependent
interactions, or by making reasonable modifications of the
in-medium nucleon-nucleon cross section.

When such discrepancies were observed in refs. \cite{Han95}, they
were attributed to the influence of the statistical decay of
fragments and other long-lived particle emitting sources. Since
the BUU model has no realistic prescription for complex fragment
formation, quantitative comparisons between BUU predictions and
experimental data require corrections taking into account properly
the contributions from such long lifetime emissions. Previous
attempts to apply such corrections could provide only a
qualitative description of measured correlations \cite{Han95}. BUU
calculations were adjusted comparing the peak values of the
calculated and measured correlation functions at $q$=20 MeV and
assuming the long-lived components to be represented as a single
lifetime secondary decay contributions. This approach did not
allow an independent test of BUU and no insights into nuclear
transport properties could be extracted. As we show below, we will
accomplish both by taking into account all the information
contained in the shape of the correlation function.

\vspace{1cm}

{\bf IMAGING ANALYSIS OF TWO-PROTON CORRELATIONS}

We study the correlation function data displayed in Fig. 1 using
the imaging analysis technique introduced in
\cite{Bro01,Bro97,Bro98}. The analysis is based on the extraction
of the source function $S(r)$ by inverting the integral equation
Eq. (3) using the measured correlation function $1+R(q)$. Thus,
"imaging the source" means inverting this equation. Such an
inversion is not completely straightforward because one has to
take into account experimental factors (limited statistics, finite
widths of momentum bins, etc.) and the intrinsic resolution of the
kernel $K(q,r)$. Small fluctuations in the data, even if well
within statistical or systematic errors, can lead to large changes
in the imaged source function. On the other hand, successful
inversion of the correlation function has the virtue of being
model independent.

The source functions $S(r)$ in our imaging analyses are expanded
in a superposition of polynomial splines \cite{Bro01},
\begin{equation}
S(r)=\sum_{j}S_{j}\cdot B_{j}(r) \label{eq08}
\end{equation}
By using this expansion, the Koonin-Pratt equation is discretized
into a matrix equation
\begin{equation}
C_{i}^{th}=1+R_{i}^{th}=\sum_{j} K_{ij}\cdot S_{j} \label{eq09}
\end{equation}
where $K_{ij}$ can be calculated from
\begin{equation}
K_{ij}=\int dr\cdot K(q,r)B_{j}(r) \label{eq10}
\end{equation}
We introduce the superscript "th" on C to distinguish theoretical
correlation functions, determined by Eq. (1) from experimental
correlation functions that we label with superscript "ex",
$C_{i}^{ex}=1+R_{i}^{ex}$.

The unknown coefficients, $S_{j}$, in the matrix equation (9), are
determined by minimizing the value of the $\chi^{2}$ between the
experimental correlation function and the one calculated from Eq.
(9):
\begin{equation}
\chi^{2}=\sum_{i}\left(C_{i}^{ex}-\sum_{j}k_{ij}S_{j}\right)^{2}/\Delta^{2}C_{i}^{ex}
\label{eq11}
\end{equation}
The conditions of the minimum can be inverted to yield an
algebraic expression for the values of $S_{j}$:
\begin{equation}
S_{j}=\sum_{i}\left[\left(K_{T}\left(\Delta^{2}C^{ex}\right)^{-1}K\right)^{-1}K^{T}B\right]_{ji}\left(C_{i}^{ex}-1\right)
\label{eq12}
\end{equation}
where $K^T$ is the transpose of the kernel matrix given by Eq.
(10). The uncertainties in the source coefficients, $\Delta S$,
are given by the square root of the diagonal elements of the
covariance matrix of the source
\begin{equation}
\Delta^{2}S=\left[K^{T}\left(\Delta^{2}C\right)^{-1}K\right]^{-1}
\label{eq13}
\end{equation}

Constraints may be included in the inversion using well defined
procedures, as described in Refs. \cite{Bro01,Bro97,Bro98}. Using
this formalism, we analyzed the data shown in Fig. 1 by
decomposing the source in a superposition of six spline
polynomials of the 3rd order over the interval 0$\leq r\leq$18 fm,
as described in Refs. \cite{Bro01} and \cite{Ver02}. The grey
areas and the hatched areas in Fig. 2 show the results of this
imaging analysis for total proton momenta $400 \leq P\leq 800$ and
$200 \leq P\leq 400$, respectively. The extracted source functions
are shown in Figs. 4-6 for E/A=80, 120 and 160 MeV, respectively.
The same grey and hatched areas as in Fig. 2 have been used to
represent the imaged sources. In Fig. 4-6, the top panel refers to
the 400$\leq P\leq$ 800 MeV/c gate, while the bottom panel refers
to the proton pairs with 200$\leq P\leq$ 400 MeV/c.

Since the imaging analysis reproduces the whole shape of the
correlation functions, the extracted sources, shown in Fig. 4-6,
provide a measure of the whole information content that can be
extracted from two-proton correlation functions. The analyses
shown in Fig. 4-6 provide significant information only for the
short-range portion of the real two-proton emitting sources,
dominated by early dynamical emissions. The $r_{1/2}$-values of
these profiles are listed in Table II and represent an estimate of
the space-time extent of these sources. Long lifetime decays
dominate mainly the tail of the source at $r\geq$10 fm. These
cannot be imaged in detail because of the poor resolution of the
data at small relative momenta ($q<$10 MeV/c). Consequently, from
the square root of the integral of these profiles, one can
estimate the fractional amount of protons emitted by the fast
dynamical source, $f_{dyn}$, by using the equation:
\begin{equation}
f_{dyn}=\sqrt{\int_{0}^{r_{max}}d^{3}r\cdot S(r)} \label{eq14}
\end{equation}
One should keep in mind that Eq. (14) gives the fraction of the
proton pairs that are emitted with $r\leq r_{max}$, and as such is
sensitive to the value of $r_{max}$. This is shown in Table II
where the $f_{dyn}$ values obtained for $r_{max}=2\cdot r_1/2$,
$r_{max}=2.5\cdot r_1/2$ and $r_{max}=3\cdot r_1/2$  are also
listed. This sensitivity issue does not influence comparisons with
BUU, because the BUU does not provide accurate predictions at
large values of r anyway. The imaging analysis does provide direct
measurements of the shape of the pre-equilibrium source at $r\leq
r_{max}$ and this can be compared directly to BUU. Only the
knowledge of how to effectively remove this sensitivity to long
time scale decay can allow quantitative comparisons between
transport theory and two-proton correlation functions. Such task
will now be accomplished for the first time. We use the imaging
technique to perform such comparisons and extract the information
contained in correlation functions regarding the transport of
nucleons during the early stages of the collision.

\vspace{1cm}

{\bf CONSTRAINING THE FRACTION OF LONG LIFETIME EMISSIONS FOR BUU
SOURCES}

The comparison of BUU simulations to experimental data can be
performed in two different ways. One can directly compare
calculated and measured correlation functions as attempted in
Figure 1, but now one can properly correct for the contributions
of long timescale decays. Alternatively, one can directly compare
the two-proton sources calculated using BUU and Eq. (4) to the
source profiles extracted from the data with the imaging
technique. We find the latter to be the more effective approach.

The key to either comparison is the correction for the influence
of long-lived decays. Essentially, the physics of these very
long-lived decays is contained in the parameter $\lambda$, which
defines the fraction of the source function that is fast and can,
in principle, be described by transport theory. Fig. 3
demonstrates the problem. The dashed line in the figure shows the
BUU source function, which is normalized so that $4\pi\int r^{2}dr
S(r)=1$, according to its definition in Eq. 4. The grey area is
the experimentally determined source function. It is normalized so
that $4\pi\int r^{2}dr S(r)=1$. However, the experimental source
must be smaller at r$<$16 fm because its normalization also
reflects the contributions at larger r from long timescale decays.
To make the BUU sources comparable to those of the data, it is
therefore necessary to likewise renormalize the BUU source,
requiring its normalization to reflect only the fraction $\lambda$
of the total source that the BUU calculations can reproduce.

To proceed further, it is imperative to correct the normalization
$\lambda$ of the BUU source function to reflect the measured
relative contributions of short-lived decays. The most accurate
normalization appears to be obtained by requiring that the
normalized source, derived from BUU
\begin{equation}
S_{dyn}^{BUU}(r)=\lambda\cdot S^{BUU}(r) \label{eq15}
\end{equation}
to most accurately replicate the measured source function by
adjusting $\lambda$ to minimize the quantity:
\begin{equation}
\chi^{2}=\sum_{i}\frac{ \left[S^{Img}(r_{i})-\lambda\cdot
S_{dyn}^{BUU}(r_{i})\right]^{2} } { \left[\Delta
S^{Img}(r_{i})\right]^{2}+\left[\lambda\cdot\Delta
  S^{BUU}(r_{i})\right]^{2}
} \label{eq16}
\end{equation}
Here, $S^{Img}(r)$ represents the imaged source, and the
uncertainties, $\Delta S^{Img}(r)$, are determined using Eq. (13).
The uncertainties of the BUU sources, $\Delta S^{BUU}(r)$, are
determined from the statistics of the simulations. The sum in Eq.
(16) is extended over the region of relative distances $r$ where
the imaged source function is still significant ($\approx$12 fm).
The BUU source, after being properly renormalized using Eq (15)
with $\lambda$=0.52, is represented in Fig. 3 by the solid line.
Clearly, one could have also obtained similar results by requiring
equality between BUU and measured values for the integral,
however, the values for this integral are extremely sensitive to
uncertainties in the measured values for $S(r)$ at large radii. We
find the normalization procedure of Eq. 15 to be more reliable
when such uncertainties are taken into consideration.

This technique has been extended to all incident energies and $P$
gates, as shown in Fig. 4-6. The imaged sources are represented by
the grey (upper panel, P=400-800 MeV/c) and the hatched (lower
panel, P=200-400 meV/c) areas. The solid, dashed and dotted lines
correspond to two-proton BUU sources for $free$, $\sigma_{in,1}$
and $\sigma_{in,0.6}$ NN cross sections, respectively. At all
incident energies, high total momentum two-proton sources are well
described by BUU simulations. Furthermore, for such energetic
protons, two-particle sources do not depend on the choice of the
N-N cross section. At low total momenta, however, the two-proton
source functions display a very strong sensitivity to details
about the in-medium N-N cross section (see lower panels in Fig.
4-6). This sensitivity is further explored in the following
section.

Before going on to the investigation of the in-medium N-N cross
section, we demonstrate in Figs. 7-9 that the renormalization of
the BUU source functions described above resolves most of the
discrepancies between measured and BUU correlation functions shown
in Fig. 1. Figure 7 shows the comparison between measured and
calculated correlation functions in collisions at E/A=80 MeV. The
left (right) panel of the figure refers to the gate 400$\leq
P\leq$ 800 MeV/c (200$\leq P\leq$ 400 MeV/c). The solid and dashed
lines correspond to the free and $\sigma_{in,0.6}$ NN cross
sections, respectively. The same conventions have been used in
Fig. 8 (E/A=120 MeV) and Fig. 9 (E/A=160 MeV). As already stated,
a comparison between the shape of the emitting sources directly
corresponds to a comparison between the shape of the correlation
peaks. Thus, it is not surprising to see that, at high total
momenta, 400$\leq P\leq$800 MeV/c (see left panels on Fig. 7-9),
BUU simulations can provide a reasonable description of the
correlation functions; this result could already be anticipated
from Fig. 4-6 (upper panels). At low total momenta (200$\leq
P\leq$ 400 MeV/c), the shape of the BUU correlations becomes very
sensitive to the details of the N-N cross section. Therefore, it
is in this low total momentum gate that two-proton correlation
analysis can provide sensitive probes of transport models of
heavy-ion collisions, provided that attention is paid to the whole
shape of the correlation functions.

\vspace{1cm}

{\bf PROBING THE NUCLEON-NUCLEON COLLISION CROSS SECTION}

The overall features of Fig. 4-6 show that for free NN cross
sections, the BUU predicts two-particle sources that are too
extended. Correspondingly, they produce two-proton correlation
peaks (Fig. 7-9), which are too narrow. This correspondence
between the size of the source and the width of the correlation
peak was already extensively illustrated in Ref. \cite{Ver02},
where a linear relation between these two quantities was derived.
At E/A=80 and 120 MeV, a reduction of the in-medium NN cross
section in BUU simulations produces sources that are in reasonable
agreement with the imaged ones. We note that this conclusion is
different from the one reached in Ref. \cite{Gon90a}. There, the
best agreement between BUU calculations and correlation function
data was attained when free cross section values were used. Our
conclusions indicate that the previous results were reached
because the influence of long-lived decays were not properly taken
into account by the analysis of Ref. \cite{Gon90a}. The results on
the cross section reductions reported here are consistent with the
conclusions deduced in Ref. \cite{Wes93} from the analysis of flow
data.

Despite these successes, at higher energies, E/A=160 MeV, the
shape of the low total momentum two-proton source from imaging
exhibits a core+tail behavior that is not reproduced by transport
model simulations. The shape of the BUU two-proton source appears
incorrect, no matter what cross section is used. However, it
should be mentioned that at E/A=160 MeV, for 200$\leq P\leq$400
MeV/c the correlation function cannot be easily normalized to 1 as
in the other cases studied in this work. The overall shape of the
correlation, if extended to larger relative momenta, suggests the
presence of space- momentum correlations that may be due to
collective motion. Such kind of correlations are different than
those described by the Koonin-Pratt formalism, and require special
study that is beyond the scope of this article. We therefore
refrain, at present, from drawing definitive conclusions about the
disagreements in the low momentum gate 200$\leq P\leq$400 MeV/c
observed in Fig. 6 and 9.

The reduction in the size of the source function when the
in-medium cross section is reduced can be simply understood. The
two-proton sources calculated from Eq. (3) and (4) correspond to
the distribution of protons when the second proton is emitted,
i.e. at the moment of its last collision with another nucleon.  As
a consequence, these sources largely reflect properties of the
phase-space distribution of particles at freeze-out when the
nucleon density is low and collisions cease. This is particularly
true when one selects low total momentum proton pairs in a range
of 200$\leq P\leq$400 MeV/c. On average, such low momentum pairs
decouple from the system at a later time when the system has
expanded further than for the high momentum pairs. We assume
schematically that the low density freeze-out stage is achieved
when the distance that a nucleon travels before colliding with
another nucleon, $d$, is of the order of the entire spatial extent
$R$ of the system  . The nucleon mean free path, $d$, can be
written as $d=1/{\rho\sigma}$, where
$\rho=A/\left(4\pi{R^{3}}/3\right)$ is the nucleon density and
$\sigma$ is the N-N cross section. It follows that one should
expect a sensitivity of the spatial extent of the source to the
N-N cross section, of the form $R\propto\sqrt{\sigma\cdot{A}}$.
This last relation explains why the size of the two-proton source
function decreases for reduced N-N cross sections, like
$\sigma_{in,1}$ and $\sigma_{in,0.6}$. The reasoning needs to be
modified, if a strong collective motion develops in the system. In
this case, the size of the sources becomes smaller than the entire
spatial extent of the system, but it can be still shown that the
direct dependence on cross section persists, with source size
$\propto\sqrt{\sigma\cdot{A}}$.

In order to confirm such qualitative arguments, we performed
additional reaction simulations reducing the cross section
alternatively when the mean local nucleon density, $\rho$, is
lower or higher that 0.5$\rho_0$ (using $\rho_{0}$=0.16 fm$^{-3}$
as the nucleon saturation density). The solid and dashed lines in
Figure 10 show the simulated temporal evolution of the central and
maximal $\rho/\rho_{0}$, respectively, during a Ar+Sc central
collision at E/A=120 MeV. (The condition of $\rho/\rho_{0}<$0.5
favors later stages of the reaction prior to the global
freeze-out.) The hatched area in Fig. 11 corresponds to the imaged
source for E/A=120 MeV and total momenta 200$\leq P\leq$400 MeV/c.
The dashed line represents the BUU two-proton source when we use a
reduced $\sigma_{in,0.6}$ NN cross-section only in the higher
density region, $\rho/\rho_{0}>$0.5, and a free NN cross section
at lower densities, $\rho/\rho_{0}<$0.5. The dot-dashed line
refers to the opposite situation, where a free NN cross section is
used for $\rho/\rho_{0}>$0.5, and $\sigma_{in,0.6}$ is used at
lower densities. The dotted and solid lines follow the same
conventions of Fig. 4-6 and correspond to sources obtained using
$\sigma_{free}$ and $\sigma_{in,0.6}$ at all densities,
respectively. It appears clearly that, by reducing the cross
section only at low density, BUU generates nearly the same
two-proton source as is generated when the cross section is
reduced at all densities. These results show that two-proton
correlation functions are sensitive to the details of the NN
collision cross sections at densities less than $\rho_{0}/2$.
According to the results of the BUU predictions shown in Fig. 10,
such low densities are typically reached during the later stages
of the reaction when the system approaches freeze-out.

\vspace{1cm}

{\bf SUMMARY AND CONCLUSIONS}

In this paper we have compared two-proton correlation functions
measured in central Ar+Sc collisions at E/A=80, 120 and 160 MeV to
BUU simulations with different choices of the NN collision cross
section. The observed discrepancies in the height of the peak at
20 MeV/c have been discussed in the context of the deficiencies of
transport theories in dealing with long-lived emitting sources due
to secondary decays of unstable fragments. The imaging technique
provides the tools to correct for such deficiencies of the model.
Indeed, this detailed shape analysis of the correlation peak
constrains the contributions from long- lived secondary decay
emissions without the need to invoke any additional BUU model
assumptions. Using these constraints, the previously observed
discrepancies between the peak values of BUU and measured
correlation functions are removed and quantitative comparisons
between theory and experiment can be made. However, the most
important consequence of the renormalization of the BUU sources
functions is that the shape of the calculated two-proton sources
in BUU can be directly compared to the shape of the imaged
sources.

Our results show that the shape of two-proton correlation
functions is very sensitive to the choice of the NN collision
cross sections in BUU simulations especially at low density. This
strong sensitivity allowed us to conclude that reduced in-medium
cross-sections are experimentally preferred. This result is
consistent with the conclusions obtained from transverse flow
analysis \cite{Wes93}. We performed an analysis of the density
dependence of such sensitivity to the NN cross section. This
analysis showed that the sensitivity the in-medium cross section
is somewhat stronger at the low densities, confirming that the
sensitivity of the correlation function to the cross-section stems
from the fact that the cross section strongly influences the
freeze-out volume.

In conclusion, this study shows that detailed imaging analyses can
provide the information needed for quantitative tests of transport
theories of heavy-ion collisions. Further investigations with
these techniques offer exciting opportunities to the influence of
collective motion on two particle correlations and to utilize the
sensitivity of two-proton correlations to further explore the
freeze-out stages of the reaction.

The National Science Foundation under Grant Nos. PHY-95-28844 and
PHY-0070818 has supported this work.

This work was performed under the auspices of the U.S. Department
of Energy by University of California, Lawrence Livermore National
Laboratory under Contract W-7405-Eng-48.

%\newpage

\newpage

\begin{table}
\caption{Used in-medium cross section (see Eq. (7)). The last
column shows the reduction of the in-medium cross section, for the
exemplary $\sigma_{free}$=30 mb, at normal nuclear density.
\label{table1}}
\begin{tabular}{ccc}
 &$\eta$&$\sigma_{in-med}/\sigma_{free}$\\
 &  &   ($\rho=\rho_{0}$) \\
\tableline
free & $\infty$ & 1 \\
$\sigma_{in,1}$ & 1.0 & 0.8 \\
$\sigma_{in,0.6}$ & 0.6 & 0.6 \\
\end{tabular}
\end{table}

\begin{table} \caption{$r_{1/2}$ and $f$ values of the imaged
sources. The $f$ values are evaluated using Eq. (14) with
$r_{max}=2{\cdot}r_{1/2}$, $r_{max}=2.5{\cdot}r_{1/2}$ and
$r_{max}=3{\cdot}r_{1/2}$. See text for more details.
\label{table2}}
\begin{tabular}{cccccc}
E/A (MeV) & P (MeV/c) & r$_{1/2} (fm)$ & f(r$_{max}$=2r$_{1/2}$) &
f(r$_{max}$=2.5r$_{1/2}$) & f(r$_{max}$=3r$_{1/2}$) \\
\tableline
80 & 200-400 & 3.76 & 0.52 & 0.59 & 0.64\\
80 & 400-800 & 3.55 & 0.62 & 0.69 & 0.73\\
\tableline
120 & 200-400 & 3.18 & 0.44 & 0.49 & 0.54\\
120 & 400-800 & 3.46 & 0.59 & 0.67 & 0.73\\
\tableline
160 & 200-400 & 2.88 & 0.48 & 0.51 & 0.54\\
160 & 400-800 & 3.38 & 0.59 & 0.66 & 0.71\\
\end{tabular}
\end{table}

\newpage

{\bf FIGURE CAPTIONS:}

Figure 1. Two-proton correlation functions in Ar+Sc central collisions at E/A=80 MeV (upper
panels), 120 MeV (central panels) and 160 MeV (bottom panels). The different data symbols
refer to two gates in total momenta, P, of the proton pairs in the CM of the reaction:
200$\leq P\leq$400 MeV/c (triangles), 400$\leq P\leq$800 MeV/c (circles).
The data are reproduced in the left and right panels.
The lines correspond to BUU simulations. The continuous lines
correspond to simulations with free NN cross sections. The dashed and dotted lines
correspond to reduced in-medium cross sections with $\eta$=1.0 and $\eta$=0.6,
respectively.

Figure 2. Two-proton correlation functions in Ar+Sc central
collisions at E/A=80 MeV (upper panels), 120 MeV (central panels)
and 160 MeV (bottom panels). Triangles: proton pairs with total
momenta 200$\leq P\leq$400 MeV/c. Circles: proton pairs with total
momenta 400$\leq P\leq$800 MeV/c. The grey and hatched areas show
the results of the imaging analysis of the experimental data.

Figure 3. Renormalization of BUU two-proton sources, as described
in Eq. 9 and 10. The grey area refers to the imaged source for
Ar+Sc reactions at E/A=80 MeV and for high total momenta, 400$\leq
P\leq$800 MeV/c. The thin dashed and solid lines represent the BUU
sources before and after renormalization, respectively.

Figure 4. The grey (upper panel) and hatched (lower panel) areas
are the two-proton sources extracted from the imaging analysis
shown in Fig. 2 for Ar+Sc central collisions at E/A=80 MeV. The
upper (lower) panel refers to a total momentum gate of 400$\leq
P\leq$800 MeV/c (200$\leq P\leq$400 MeV/c). The thin solid, dotted
and dashed lines represent the renormalized BUU sources using
free, $\sigma_{in,1}$ and $\sigma_{in,0.6}$ NN collision cross
sections, respectively.

Figure 5. The grey (upper panel) and hatched (lower panel) areas
are the two-proton sources extracted from the imaging analysis
shown in Fig. 2 for Ar+Sc central collisions at E/A=120 MeV. The
upper (lower) panel refers to a total momentum gate of 400$\leq
P\leq$800 MeV/c (200$\leq{P}\leq$400 MeV/c). The thin solid,
dotted and dashed lines represent the renormalized BUU sources
using free, $\sigma_{in,1}$ and $\sigma_{in,0.6}$ NN collision
cross sections, respectively.

Figure 6. The grey (upper panel) and hatched (lower panel) areas
are the two-proton sources extracted from the imaging analysis
shown in Fig. 2 for Ar+Sc central collisions at E/A=160 MeV. The
upper (lower) panel refers to a total momentum gate of 400$\leq
P\leq$800 MeV/c (200$\leq P\leq$400 MeV/c). The thin solid, dotted
and dashed lines represent the renormalized BUU sources using
free, $\sigma_{in,1}$ and $\sigma_{in,0.6}$ NN collision cross
sections, respectively.

Figure 7. The data points show measured two-proton
correlations in central Ar+Sc reactions at E/A=80 MeV.
The left (right) panel corresponds to a total momentum
gate 400$\leq P\leq$800 MeV/c (200$\leq P\leq$400 MeV/c).
The thin solid and dashed lines refer to BUU correlation
functions after the BUU two-proton sources have been
renormalized as described in the text and in Fig. 3.

Figure 8. The data points show measured two-proton
correlations in central Ar+Sc reactions at E/A=120 MeV.
The left (right) panel corresponds to a total momentum
gate 400$\leq P\leq$800 MeV/c (200$\leq P\leq$400 MeV/c).
The thin solid and dashed lines refer to BUU correlation
functions after the BUU two-proton sources have been
renormalized as described in the text and in Fig. 3.

Figure 9. The data points show measured two-proton
correlations in central Ar+Sc reactions at E/A=160 MeV.
The left (right) panel corresponds to a total momentum
gate 400$\leq P\leq$800 MeV/c (200$\leq P\leq$400 MeV/c).
The thin solid and dashed lines refer to BUU correlation
functions after the BUU two-proton sources have been
renormalized as described in the text and in Fig. 3.

Figure 10. Central (solid line) and maximal (dashed line) nuclear matter density $\rho$,
normalized to saturation density, $\rho_{0}$ (= 0.17 fm-3), as a function of time,
$t$, during a central Ar+Sc collision at E/A=120 MeV, as simulated by the BUU model.

Figure 11. Hatched area: two-proton source function extracted from
imaging analysis of Ar+Sc collisions at E/A=120 MeV and for low
total momentum proton pairs (200$\leq P\leq$400 MeV/c). The solid
and dotted lines correspond to two-proton sources simulated by
BUU, after renormalization, and using free and $\sigma_{in,0.6}$
NN cross sections, respectively. The dashed and dot-dashed line
refer to simulations where NN cross sections have been reduced
only at higher densities ($\rho/\rho_{0}>0.5$) or only at lower
densities ($\rho/\rho_{0}<0.5$), respectively.

\end{document}